\documentclass[aoas,MSNbibl,nameyear,dvips]{arximspdf}
\usepackage{dcolumn}
\usepackage{graphicx}

\doi{10.1214/11-AOAS468}
\volume{5}
\issue{3}
\pubyear{2011}
\firstpage{1920}
\lastpage{1947}

\makeatletter
\newcolumntype{d}[1]{D{.}{.}{#1}}
\makeatother

\begin{document}
\begin{frontmatter}

\title{Spatial modeling of the 3D morphology of hybrid polymer-ZnO solar cells,
    based on electron tomography data\thanksref{T1}}
\runtitle{Modeling of the 3D morphology of polymer solar cells}
\thankstext{T1}{Supported by
    Deutsche Forschungsgemeinschaft (DFG) under the Priority
    Programme: ``Elementary Processes of Organic Photovoltaics'' (SPP 1355).}

\begin{aug}
\author[A]{\fnms{O.} \snm{Stenzel}\corref{}\ead[label=e1]{ole.stenzel@uni-ulm.de}},
\author[A]{\fnms{H.} \snm{Hassfeld}\ead[label=e2]{henrik.hassfeld@uni-ulm.de}},
\author[A]{\fnms{R.} \snm{Thiedmann}\ead[label=e3]{ralf.thiedmann@uni-ulm.de}},
\author[B]{\fnms{L. J. A.} \snm{Koster}\ead[label=e4]{l.j.a.koster@tue.nl}},
\author[B]{\fnms{S.~D.}~\snm{Oosterhout}\ead[label=e5]{s.d.oosterhout@tue.nl}},
\author[B]{\fnms{S. S.} \snm{van Bavel}\ead[label=e6]{s.s.v.bavel@tue.nl}},
\author[B]{\fnms{M. M.} \snm{Wienk}\ead[label=e7]{m.m.wienk@tue.nl}},
\author[B]{\fnms{J.} \snm{Loos}\ead[label=e8]{j.loos@tue.nl}},
\author[B]{\fnms{R.~A.~J.}~\snm{Janssen}\ead[label=e9]{r.a.j.janssen@tue.nl}}
and
\author[A]{\fnms{V.} \snm{Schmidt}\ead[label=e10]{volker.schmidt@uni-ulm.de}}

\runauthor{O. Stenzel et. al.}

\affiliation{Ulm University, Ulm University, Ulm University, Eindhoven
University of Technology,
Eindhoven University of Technology, Eindhoven University of Technology,
Eindhoven University of Technology,
Eindhoven University of Technology, Eindhoven University of Technology
and Ulm University}

\address[A]{O. Stenzel\\
H. Hassfeld\\
R. Thiedmann\\
V. Schmidt\\
Institute of Stochastics\\
Ulm University\\
89069 Ulm\\
Germany \\
\printead{e1}\\
\hphantom{E-mail: }\printead*{e2}\\
\hphantom{E-mail: }\printead*{e3}\\
\hphantom{E-mail: }\printead*{e10}}

\address[B]{L. J. A. Koster\\
S. D. Oosterhout\\
S. S. van Bavel\\
M. M. Wienk\\
J. Loos\\
R. A. J. Janssen\\
Chemical Engineering and Chemistry\\
Molecular Materials and Nanosystems\\
Eindhoven University of Technology\\
5600 MB Eindhoven\\
The Netherlands\\
\printead{e4}\\
\hphantom{E-mail: }\printead*{e5}\\
\hphantom{E-mail: }\printead*{e6}\\
\hphantom{E-mail: }\printead*{e7}\\
\hphantom{E-mail: }\printead*{e8}\\
\hphantom{E-mail: }\printead*{e9}}
\end{aug}

\received{\smonth{8} \syear{2010}}
\revised{\smonth{3} \syear{2011}}

%
\begin{abstract}
A spatial stochastic model is developed which describes the 3D
nanomorphology of composite materials, being blends of two
different (organic and inorganic) solid phases. Such materials are
used, for example, in photoactive layers
of hybrid polymer zinc oxide solar cells. The model is based on ideas
from stochastic geometry and spatial statistics.
Its parameters are fitted to image data gained by electron
tomography (ET), where adaptive thresholding and stochastic
segmentation have been used to represent morphological features
of the considered ET data by unions of overlapping spheres. Their
midpoints are modeled by a stack of 2D point processes with a
suitably chosen correlation structure, whereas a moving-average
procedure is used to add the radii of spheres. The model is
validated by comparing physically relevant characteristics of real
and simulated data, like the efficiency of exciton quenching,
which is important for the generation of charges and their
transport toward the electrodes.
\end{abstract}

%

\end{frontmatter}

\section{Introduction}\label{sec.Introduction}

Using methods from stochastic geometry and spatial statistics, a
stochastic model is developed which describes the 3D
nanomorphology of composite materials, being blends of two
different (organic and inorganic) solid phases. Such materials are
used, for example, in photoactive layers of hybrid polymer zinc oxide (ZnO)
solar cells where the two solid phases play the role of a
polymeric electron donor, consisting of, for example,
poly(3-hexylthiophene), and an inorganic ZnO-electron acceptor,
respectively. There is a great advantage of polymer solar cells
due to their potentially low production costs, in comparison with
classical silicon solar cells. However, the efficiency of polymer
solar cells critically depends on the intimacy of mixing of the
donor and acceptor semiconductors used in these devices to create
charges as well as on the presence of unhindered percolation
pathways in the individual solid phases of the composite material
to transport positive and negative charges toward electrodes; see, for
example, \citet{Yang07}. It is therefore very important to have tools
at one's disposal which are suitable to analyze and model the 3D
morphology of these materials quantitatively. So far, no such
tools are available in literature due to the fact that imaging of
the 3D morphology in high resolution is a difficult task. The
first 3D images of photoactive layers in polymer solar cells,
gained by means of electron tomography (ET), have been published
only recently; see \citet{Bavel09} and \citet{Oosterhout09}.

In the present paper, such 3D images are used to fit
our model to real data. The model then helps to get a better insight
into the impact of morphology on the performance of
polymer solar cells and, simultaneously, it can be used for virtual
scenario analyses, where model-based morphologies of solar cells
are simulated to identify polymer solar cells with improved
nanostructures.

The model developed in this paper is based on methods from
stochastic geo\-metry and spatial statistics; see \citet{KEN10}
and \citet{Gelfand10} for comprehensive surveys on recent results
in these fields. In particular, stationary marked point processes
are considered as models for complex point patterns extracted from
ET images, where the points are associated with additional
information, so-called ``marks.''

Note that point processes in 3D have been used for many years to
analyze geometrically complex point patterns; see, for example,
\citet{BAD87}. More recently, further case studies in 3D point
process modeling have been performed, for example, in \citet{BDS05},
\citet{BFPS05} and \citet{SGM05}; see also \citet
{BAD06}. Besides,
there are many monographs dealing with point processes in
multidimensional spaces and their statistical inference and
simulation. We refer, for instance, to \citet{DAL08},
\citet{DIG03}, \citet{Illian08}, \citet{MOE04}, as well as
\citet{Stoyan95}.

The paper is organized as follows. Section~\ref{sec.pre.lim}
briefly describes the considered solar cells and the corresponding
image data on which the model is based.
In particular, in Section~\ref{sec.multi-scale}, the main ideas of
a multi-scale approach to the segmentation of 3D images are
summarized, which has been developed recently in \citet{TH10}. The
crucial step of this approach is to find an efficient
representation of the binarized and morphologically smoothed
images by unions of overlapping spheres.\vadjust{\eject}

Then, in Section~\ref{sec.SimulationModel}, the spatial stochastic
model for the ZnO phase is introduced, separately for
morphologically smoothed ZnO domains (macro-scale) and for those
parts representing the difference between the smoothed and
nonsmoothed binary images (micro-scale). Based on unions of
overlapping spheres representing the ZnO domains, that is, marked
point patterns extracted from ET images, a stochastic model is
built for the smoothed 3D morphology (macro-scale) of the
photoactive layers considered in this paper. Since a strong
correlation of midpoints of spheres in $z$-direction is observed,
we propose a multi-layer approach considering sequences of
correlated 2D point processes to model the 3D point patterns of
midpoints. The members of these sequences belong to a suitably
chosen class of planar Poisson cluster processes, being parallel
to the $x$--$y$-plane. In particular, a generalized version of
Mat{\'e}rn cluster processes is considered, where the cluster
points are scattered in uniformly oriented ellipses around their
cluster centers (Section~\ref{sec.mod.sli}). To model the 3D point
patterns of midpoints, a Markov chain with stationary initial
distribution is constructed, which consists of highly correlated
Mat{\'e}rn cluster processes (Section~\ref{sec.spa.bir}). It can
be seen as a stationary point process in 3D, where the radii of
spheres are considered as marks of this point process
(Section~\ref{sec.Sim.rad}). Subsequently, a spatial stochastic
model for the micro-scale part of the morphological structure is
developed (Section~\ref{sec.SimulationModelMicro}). It is used to
invert morphological smoothing and completes our model for the 3D
morphology of hybrid polymer-ZnO solar cells. Furthermore, a
method to fit model parameters to real image data is
proposed.\looseness=-1

Section~\ref{sec.mod.val} deals with model validation. To evaluate
the goodness of fit, we compare model characteristics which have
been computed from real and simulated data, respectively, like the
volume fractions of voxels contributing to monotonous percolation
pathways through the photoactive layer, the distribution of
spherical contact distances, and the probabilities of exciton
quenching. These characteristics have already been considered in
\citet{Oosterhout09}, since they are strongly related with the
performance of solar cells. Finally, Section~\ref{sec.con.out}
concludes and provides a short outlook regarding possible future
research.

\section{Polymer solar cells}\label{sec.pre.lim}

In this section some necessary background information regarding
the functionality of polymer solar cells is provided, together
with corresponding image data on which the model is based.

\setcounter{footnote}{1}

\subsection{Photoactive layers}\label{sub.pho.act} 
\begin{figure}

\includegraphics{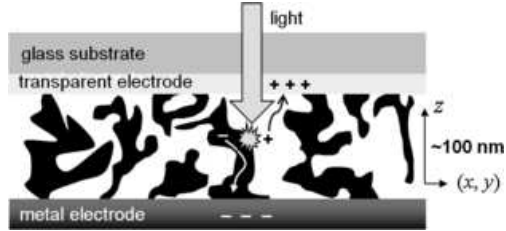}
\caption{Schematic layout of a polymer-ZnO thin
film solar cell, showing the percolation of photogenerated holes ($+$)
and electrons ($-$).}\label{fig.schematicView}
\end{figure}

We consider
photoactive layers of hybrid polymer zinc oxide (ZnO) solar cells
where the two solid phases play the role of a~polymeric electron
donor, consisting of, for example, poly(3-hexylthio\-phene), and an inorganic
ZnO-electron acceptor, respectively. Upon exposure to light,
photons are absorbed in the polymer phase and so-called
``excitons,'' that is, photoexcited electron--hole pairs, evolve.
Excitons are neutral quasi-particles which diffuse inside the
polymer phase within a limited lifetime; see \citet{Sha08}. If an
exciton reaches the interface to the ZnO phase, it is split up
into a free electron (negative charge) in the ZnO and a hole
(positive charge) in the polymer phase. This process is commonly
referred to as quenching, because it reduces the intrinsic
fluorescent decay of the exciton in the polymer. Provided that the
electrons in the ZnO phase and the holes in the polymer phase
reach the electrodes at the top and bottom of the photoactive
layer, respectively, photocurrent is generated. A schematic
illustration of the morphology of photoactive layers in hybrid
polymer-ZnO solar cells is shown in
Figure~\ref{fig.schematicView}, where the electrodes are supposed
to be parallel to the $x$--$y$-plane. For further information
about polymer solar cells and the physical processes therein we
refer, for example, to \citet{Bra08}.

Note that the extent of blending of the two materials has a
large impact on the efficiency of these solar cells, because not
all excitons are quenched due to their limited lifetimes. Thus, a
morphology as displayed in Figure~\ref{fig.schematicView}, where
both materials are mixed intimately, is desirable since the
excitons are likely to reach the interface and charges can be
generated. In other words, for a morphology which would be ideal
with respect to this aspect of functionality, each location of the
polymer phase should have a distance to the ZnO phase that is
smaller than the diffusion length of excitons. For each location
within the polymer phase, the fraction of excitons reaching the
interface is called the quenching probability at this location.
The mean of these quenching probabilities, that is, the quenching
probability at a randomly chosen location of the polymer phase, is
called the quenching efficiency.

Furthermore, the existence of unhindered percolation pathways
within both phases, ZnO and polymer, is crucial since the
generated charges have to be transported to the electrodes
throughout the phases. Because of the electric field between the
electrodes, these pathways should be preferably monotonous. Hence,
to obtain solar cells with high efficiency, an intimately mixed
morphology with monotonous percolation pathways for both charge
carriers is desirable and should be taken into account when
producing devices. The stochastic model developed in the present
paper will be used to identify morphologies with improved\vadjust{\eject}
efficiency by generating virtual morphologies and investigating
the transport processes of electrons and excitons, respectively.
This will be the subject of a forthcoming paper.

\subsection{Electron tomography images}\label{sub.tom.gra}

The image data have been gained by electron tomography (ET); see
\citet{Bavel09} and \citet{Oosterhout09}. In particular, we consider
images for three devices with different thicknesses of the
photoactive layers: 57~nm, 100~nm and 167~nm.
For each of the three thicknesses, the 3D ET images are given as
stacks of 2D grayscale images (being parallel to the
$x$--$y$-plane, say), which are numbered according to their
location in $z$-direction. The sizes of these images in the
$x$--$y$-plane are $934\times911$ voxels for the 57~nm film, and
$942\times911$ voxels for the other two thicknesses. Each voxel
represents a cube with side length of 0.71~nm.

\begin{figure}[b]

\vspace*{-2pt}
\includegraphics{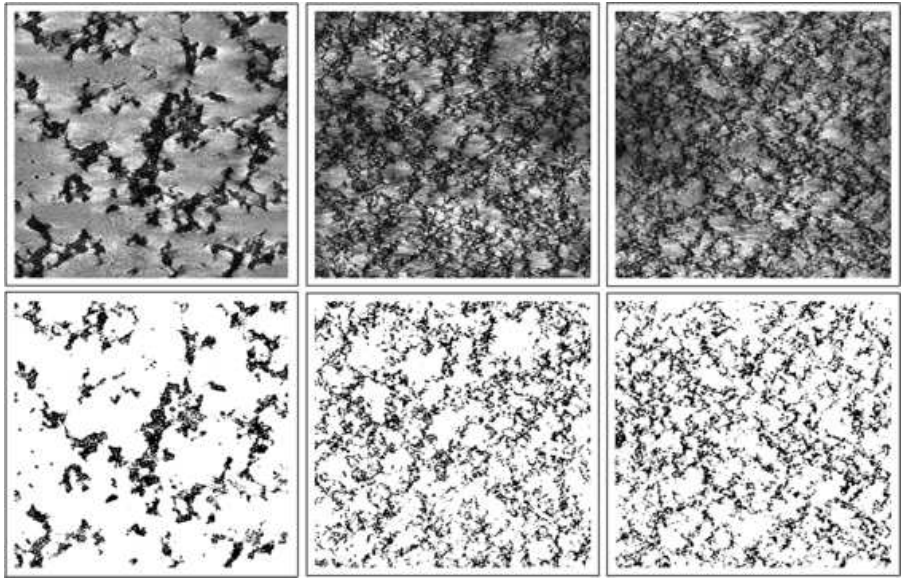}

\vspace*{-2pt}
\caption{2D images of hybrid polymer-ZnO solar cells;
first column: 57~nm, second: 100~nm, third: 167~nm; first row:
original grayscale images, second:
binarized images.}\label{fig.raw_data_slices}
\end{figure}

Figure~\ref{fig.raw_data_slices} shows representative 2D slices
for the three film thicknesses, where the darker parts of the
images represent the ZnO phase due to a higher electron density
compared to polymer. The images displayed in
Figure~\ref{fig.raw_data_slices} indicate clear structural
differences for the three films. With increasing layer thickness,
the separated domains of polymer and ZnO are getting finer. In
particular, the thinnest film, that is, the photoactive layer with
thickness of 57~nm, features large domains of both polymer and
ZnO. The stochastic 3D model to be fitted takes these
morphological differences into account. More precisely, the model
type will be the same for all three layer thicknesses. Only the
values of some model parameters will be different for the varying
morphologies; see Section~\ref{sec.SimulationModel} below.

To develop a stochastic model for the 3D morphology of hybrid
polymer-ZnO solar cells, the 3D ET grayscale images have to be
binarized appropriately. Binarization is necessary since we need
to decide which voxels are classified as polymer and which as ZnO.
An elementary approach to binarize grayscale images is to use a
global threshold: voxels are set to white (polymer) if their
grayscale value exceeds a certain threshold, and are otherwise set
to black (ZnO). However, it is difficult to find a single global
threshold to binarize the ET images because of the irregular
brightness of these images. Thus, instead of considering global
thresholds, a~method of adaptive thresholding has been used for
binarization; see \citet{TH10}. This method is based on techniques
of \citet{Yanowitz89} and \citet{Blayvas06}, where the main idea
is to construct a threshold surface which is location-dependent
and takes local conditions like overexposure or underexposure into
account. Examples of binarizing the ET images by adaptive
thresholding are displayed in Figure~\ref{fig.raw_data_slices}.

\subsection{Segmentation of binarized images}\label{sec.multi-scale}

In this section we briefly summarize the main ideas of a
multi-scale approach to the segmentation of 3D images, which has
been developed recently in \citet{TH10}. The crucial step is to
find an efficient representation of the binarized and
morphologically smoothed images by unions of overlapping spheres.

\begin{figure}[t]

\includegraphics{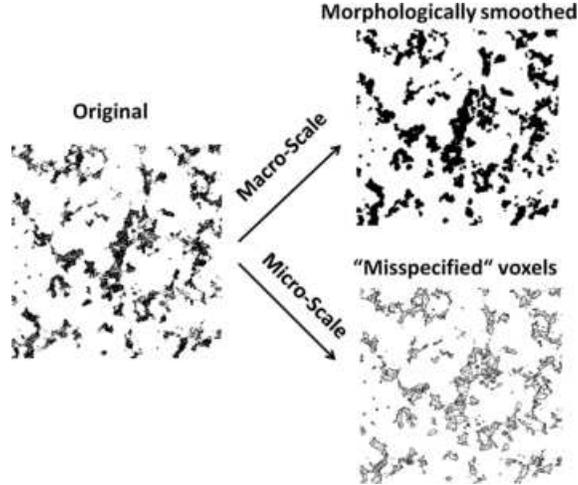}

\vspace*{-2pt}
\caption{Original image split up into structural components at two
different scales.}\label{fig:MicroMacroScale}
\vspace*{-2pt}
\end{figure}

Let $B$ denote the ZnO phase of the binarized images. Since the
morphology of the set $B$ is rather complex (see
Figure~\ref{fig.raw_data_slices}), it is difficult to describe
this morphology directly, just by a single stochastic model. We
therefore developed a multi-scale approach to represent the ZnO
phase by different structural components. Each of them will be
described separately by suitably chosen stochastic models. More
precisely, we distinguish between a macro-scale component of the
binarized ET images, which is obtained by morphological smoothing,
and several micro-scale components, which consist of those voxels
that have been misspecified by the morphological smoothing; see
Figure~\ref{fig:MicroMacroScale}. The intention of morphological
smoothing is to reduce the structural complexity of the binarized
ET images, that is, to omit very fine structural components such as
``thin ZnO branches,'' that is, thin ZnO parts connected to larger ZnO
domains, ``isolated ZnO particles,'' that is, small ZnO particles in
the polymer domains, and ``polymeric holes,'' that is, small polymeric
particles inside the ZnO domains. The morphological
transformations which we use for smoothing the ZnO phase $B$ of
the original binarized ET images are twofold: ``dilation'' and
``erosion.'' The morphologically smoothed version of the set $B$
will thus be denoted by $B^{\prime\prime}$.

In the next step a stochastic algorithm [see \citet{TH10}] is used
to efficiently represent the set $B^{\prime\prime}$ by a union of
spheres, which is denoted by $B^{\prime\prime\prime}$. This leads
to an enormous data reduction. Another advantage of this
representation of the set $B^{\prime\prime}$ by unions of spheres
is that it allows the interpretation of the morphologically
smoothed ZnO phase as a realization of a marked point process
where the midpoints of the spheres are the points and the
corresponding radii the marks.

\begin{figure}[b]

\vspace*{-2pt}
\includegraphics{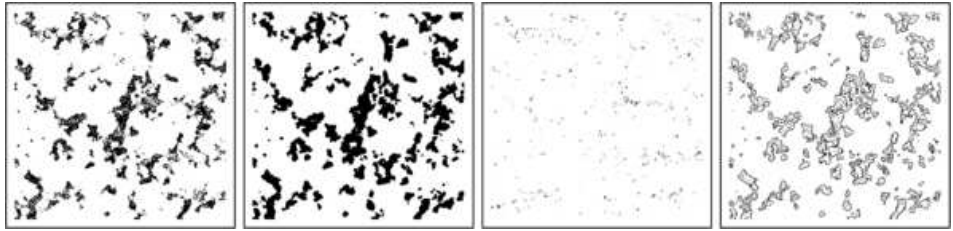}%
\vspace*{-2pt}

\caption{First: binarized (nonsmoothed) 2D slice of 57~nm file;
second: representation by unions of spheres; third: outer
misspecifications; fourth: inner misspecifications (boundary
and interior).}\label{fig.misspecification}
\end{figure}

Recall that in the macro-scale component $B^{\prime\prime}$ of the
binarized ET images as well as in its representation
$B^{\prime\prime\prime}$ by unions of spheres, some structural
details of the original ZnO phase $B$, like isolated particles,
thin branches and polymeric holes, are omitted. Furthermore, the
boundaries of ZnO domains are morphologically smoothed and
slightly enlarged by dilation. Hence, when comparing the sets $B$
and $B^{\prime\prime\prime}$, we observe that some voxels are
misspecified, that is, indicated as ZnO although originally being
polymer, and vice versa. The set $B\triangle
B^{\prime\prime\prime}= (B\cup B^{\prime\prime\prime})\setminus
(B\cap B^{\prime\prime\prime})$ of misspecified voxels is
subdivided into several subcomponents, where each of these
subcomponents will be modeled separately. First, two main types
of misspecifications are distinguished: outer misspecifications
and inner misspecifications; see
Figure~\ref{fig.misspecification}. Each ZnO voxel that is not
covered by a sphere, that is, belonging to the set $B\setminus
B^{\prime\prime\prime}$ and therefore
constituted as polymer, is called an outer misspecification.
Typically, thin branches and isolated ZnO particles
are outer misspecifications. On the other hand, each polymer voxel
which is covered by a sphere,
that is, belonging to $B^c\cap B^{\prime\prime\prime}$ and constituted
as ZnO, is called an inner misspecification.
Inner misspecifications are further subdivided
into boundary misspecifications and interior misspecifications. On the
one hand, polymer voxels (belonging to $B^c$), located near
the boundary $\partial B^{\prime\prime}$ of the macro-scale component
$B^{\prime\prime}$
and covered by a sphere, are called boundary misspecifications.
On the other hand, each inner misspecification which is not a boundary
misspecification is called an interior misspecification.
Typically, polymeric holes belong to interior misspecifications.

\section{Stochastic modeling}\label{sec.SimulationModel}

We now present our approach to stochastic modeling of the 3D
nanomorphology of the ZnO phase in photoactive layers with three
different thicknesses which are given by the binarized ET images
described in Section~\ref{sub.tom.gra}. Note that the model type
is the same for all three thicknesses, just the fitted values of
some model parameters are different. This means, in particular, that
our model can be used for computer-based scenario analyses with
the general objective of developing improved materials and
technologies for polymer solar cells.

In accordance with the multi-scale representation of the ZnO phase
which has been described in Section~\ref{sec.multi-scale}, we will
establish stochastic simulation models separately for the
morphologically smoothed ZnO domains represented by unions of
overlapping spheres, that is, the macro-scale representation of the
ZnO phase, and for the three types of misspecifications, that is, the
micro-scale components.

\subsection{Point-process model for systems of overlapping
spheres}\label{sec.SimulationModelMacro}

To begin with, we develop a point-process model which describes the
macro-scale component of the ZnO phase represented by unions of
spheres. This model is constructed in several steps. First we
consider 2D point processes for those midpoints of spheres which
belong to single slices of voxels, being parallel to the
$x$--$y$-plane. Since a strong correlation of midpoints spheres in
$z$-direction is observed, we propose a multi-layer approach
considering sequences of correlated 2D point processes to model
the 3D point patterns of midpoints.\looseness=-1

The members of these sequences belong to a suitably chosen class
of planar Poisson cluster processes. In particular, elliptical
Mat{\'e}rn cluster processes are considered, where the cluster
points are scattered in ellipses of uniformly distributed
orientation around their cluster centers. To model the 3D point
patterns of midpoints, a Markov chain with stationary initial
distribution is constructed, which consists of highly correlated
Mat{\'e}rn cluster processes. It can be seen as a stationary point
process in 3D, where the radii of spheres are considered as
marks of this point process. The mark correlation functions, which have\vadjust{\eject}
been computed for the radii of spheres extracted from ET images,
show strong positive correlations for small distances between
midpoints. Hence, for a given configuration of midpoints, the
radii associated with these midpoints are not modeled just by
independent marking, but a certain moving-average procedure is
proposed. Note that our model of a stationary marked point
process in 3D describing the macro-scale component of the ZnO
phase is not isotropic. This is in accordance with the
nanomorphology observed in the ET images; see
\citet{Oosterhout09}.

\subsubsection{Elliptical Mat{\'e}rn cluster processes}\label{sec.mod.sli}

\begin{figure}[b]

\vspace*{-2pt}
\includegraphics{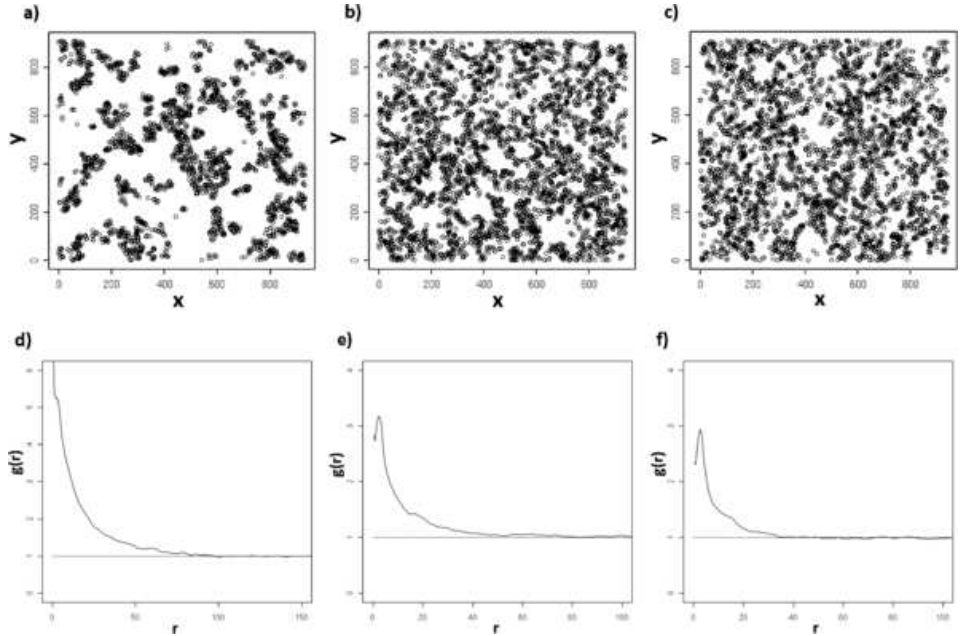}%
\vspace*{-2pt}
\caption{Top: point patterns of midpoints in 2D slices ($\mathrm{a} = 57$ nm,
$\mathrm{b}= 100$ nm, $\mathrm{c}= 167$~nm);
bottom: estimated pair correlation functions ($\mathrm{d} = 57$ nm, $\mathrm{e} = 100$
nm, $\mathrm{f} = 167$~nm).}\label{fig.cor.fun}
\end{figure}

To get an idea which class of point processes is suitable to model
the midpoints belonging to the individual slices of voxels, we consider
the pair correlation function $g\dvtx(0,\infty)\to(0,\infty)$\vadjust{\goodbreak}
of stationary and isotropic point processes in $\mathbb{R}^2$. Note that
$g(r)$ is proportional to the relative frequency of point pairs with
distance $r>0$ from
each other; see, for example, \citet{Illian08}.
Then, for each of the three photoactive
layers with thicknesses of 57~nm, 100~nm, and 167~nm, the values~$\widehat g(r)$ of
the pair correlation function have been estimated for distances $r$
within some interval $(0,r_{\max})$; see Figure~\ref{fig.cor.fun}.

Since $\widehat g(r)>1$ for small $r>0$, these estimates clearly
indicate clustering of points, which can also directly be seen from
the point patterns shown in Figure~\ref{fig.cor.fun}. The clusters
appearing in these point patterns\vadjust{\eject} seem to be located in
relatively small (bounded) areas, which corresponds to the fact that
$\widehat g(r)\approx1$ for sufficiently large $r>0$.
The shapes of the clusters are not circular, but
rather elongated. Hence, we propose to consider Mat{\'e}rn cluster processes,
where the cluster points are scattered in ellipses of uniformly
distributed orientation around their cluster centers.

\begin{figure}[b]
\includegraphics{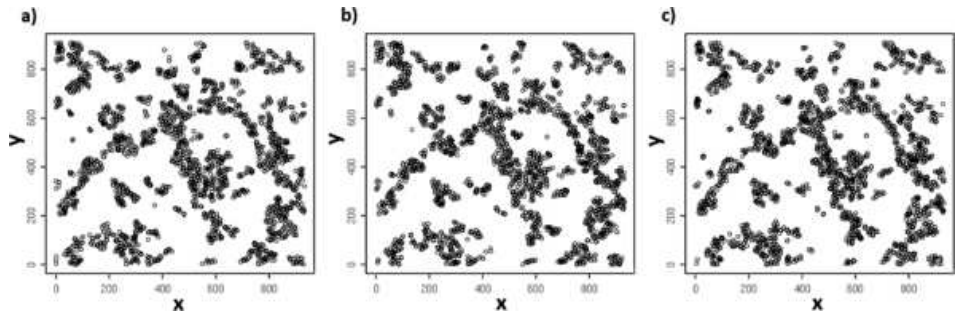}%
\caption{Point patterns of midpoints for successive 2D-slices from the
57~nm data set
($\mathrm{a} = \mbox{slice }35$, $\mathrm{b} = \mbox{slice }36$, $\mathrm{c} = \mbox{slice
}37$).}\label{fig.SuccessiveMidpoints}
\end{figure}

This class of (elliptical) Mat\'ern cluster processes in
$\mathbb{R}^2$ can be described by a vector of four parameters:
$(\lambda_c,\lambda_d,a,b)$, where $\lambda_c$ is the intensity of
the stationary Poisson point process $\{T_n, n\ge1\}$ of cluster
centers, $a$ and $b$ with $a>b>0$ are the semi-axes of (random)
ellipses $\mathcal{E}_{a,b}(T_n,\zeta_n) \subset\mathbb{R}^2$
centered at the points $T_n$ of the Poisson process $\{T_n\}$ of
cluster centers and rotated around $T_n$ by random angles
$\zeta_n$ which are independent and uniformly distributed on the
interval $[0,\pi)$, and $\lambda_d$ is the intensity of the
stationary Poisson processes $\{S_{ni}, i\ge1\}$ of cluster
members which are released by the cluster centers $T_n$ within the
ellipses $\mathcal{E}_{a,b}(T_n,\zeta_n)$. The Mat\'ern cluster
process is then defined as the random point pattern $\{S_n\}$
given by $\{S_n\}= \bigcup_{n=1}^\infty
(\{S_{ni}, i\ge1\}\cap
\mathcal{E}_{a,b}(T_n,\zeta_n) )$, where the sequences
$\{\zeta_n\}, \{T_n\}, \{S_{1i}\}, \{S_{2i}\}, \ldots$ are
assumed to be independent.

\subsubsection{Markov chain of Mat{\'e}rn cluster processes}\label{sec.spa.bir}

There are strong simila\-rities between consecutive 2D
slices in terms of a high correlation of midpoint locations in
$z$-direction as well as approximately equal numbers of points.
Figure~\ref{fig.SuccessiveMidpoints} shows a series of such
consecutive 2D slices from the 57~nm data~set.

As a consequence, it is not suitable to model the stacks of these 2D
point patterns by sequences of independent Mat{\'e}rn processes.
But the (vertical) correlation structure visualized in Figure~\ref
{fig.SuccessiveMidpoints} can be taken
into account by considering a Markov chain of Mat{\'e}rn processes.
This allows us to model small displacements of clusters when passing
from slice to slice. Furthermore, ``births'' and
``deaths'' of clusters in $z$-direction can also be modeled in this
way. In other words, we consider a certain class of spatial
birth-and-death processes with random displacement of points; see, for
example, \citet{MOE04}.\vadjust{\eject}

For each integer $z\ge1$, let $\{B^{(z)}_n, n\ge1\}$ be a stationary
Poisson point process in $\mathbb{R}^2$ with intensity $\lambda_c^\prime$
such that $0<\lambda_c^\prime<\lambda_c$,
and let $\{ \delta^{(z)}_n, n\ge1\}$ be
an independent and identically distributed (i.i.d.) sequence of Bernoulli
random variables, which is independent of $\{B^{(z)}_n\}$, where
$P(\delta^{(z)}_n=1)=p$ for some $p\in(0,1)$. Note that
$\{B^{(z)}_n\}$ and $\{\delta^{(z)}_n\}$ will be used in order to model
``births'' and ``deaths'' of cluster centers, respectively.

For each integer $z\ge1$, an i.i.d. sequence $\{D_n^{(z)}\}$ of random
displacement vectors
$D_1^{(z)},D_2^{(z)},\ldots$ with values in $\mathbb{R}^2$ is
considered, which is independent of $\{B^{(z)}_n\}$
and $\{ \delta^{(z)}_n\}$. We assume that the random vectors
$D_1^{(z)},D_2^{(z)},\ldots$ are uniformly distributed in
the set $b(o,r^{\prime\prime})\setminus b(o,r^\prime)$, where $r^\prime
$ and $r^{\prime\prime}$ denote the size of minimum and maximum
displacement, respectively;
$0<r^\prime<r^{\prime\prime}$.

\begin{figure}[b]
\vspace*{5pt}
\includegraphics{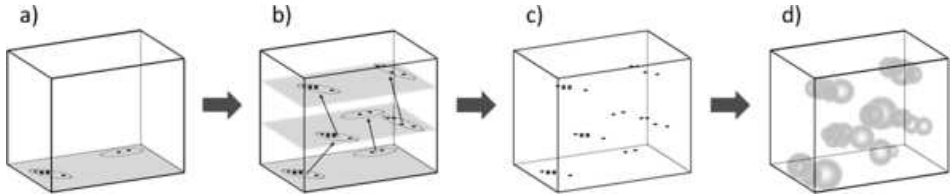}

\caption{\textup{(a)} initial 2D point pattern; \textup{(b)} displacement of cluster
centers, including ``birth'' and
``death;''
\textup{(c)} 3D point pattern; \textup{(d)} union of spheres.}\label{fig.3DStack}
\end{figure}

Then, a (stationary) Markov chain $\{\{S_n^{(z)}\}, z\ge1\}$ of
Mat{\'e}rn processes can be constructed as follows. For $z=1$, let
$\{S_n^{(1)}\}$ be an elliptical Mat{\'e}rn cluster process as
introduced in Section~\ref{sec.mod.sli}, that is, $\{S^{(1)}_n\}=
\bigcup_{n=1}^\infty (\{S_{ni}, i\ge1\}\cap
\mathcal{E}_{a,b}(T_n,\zeta_n) )$. Assume that the ``birth
rate'' $\lambda_c^\prime$ and the ``survival probabili\-ty''~$p$
satisfy $\lambda_c p+\lambda_c^\prime=\lambda_c$, where
$\lambda_c$ is the intensity of the Poisson process $\{T_n\}$ of
cluster centers. For $z=2$, the Poisson process $\{T^{(2)}_n\}$ of
cluster~cen\-ters is then given by\vspace*{-3pt}
$\{T^{(2)}_n, n\ge1\}=\bigcup_{j\dvtx
\delta_j^{(1)}=1}\{T_j^{(1)}+D_j^{(1)}\} \cup
\{B^{(1)}_n, n\ge1 \}$. The Poisson process
$\{T^{(3)}_n\}$ is constructed in the same way as $\{T^{(2)}_n\}$,
that is,
\[
\bigl\{T^{(3)}_n, n\ge1\bigr\}=\bigcup_{j\dvtx\delta_j^{(2)}=1}\bigl\{
T^{(2)}_j+D_j^{(2)}\bigr\}
\cup \bigl\{B^{(2)}_n, n\ge1 \bigr\} ,
\]
and so on; see also Figure~\ref{fig.3DStack}. The Mat{\'e}rn
processes $\{S_n^{(2)}\},\{S_n^{(3)}\},\ldots$ are built similarly
to the construction of $\{T^{(2)}_n\},\{T^{(3)}_n\},\ldots.$ For
example, $\{S_n^{(2)}\}$ is given by
\begin{eqnarray*}
\bigl\{S^{(2)}_n\bigr\} &=& \bigcup_{j\dvtx\delta_j^{(1)}=1} \bigl(\bigl\{S_{ji}+D_j^{(1)}
, i\ge1\bigr\}\cap\mathcal{E}_{a,b}\bigl(T_j^{(1)}+D_j^{(1)},\zeta_j\bigr) \bigr)\\
&&{}
\cup\bigcup_{n=1}^\infty \bigl(\bigl\{S^{(1)}_{ni}, i\ge1\bigr\}\cap\mathcal
{E}_{a,b}\bigl(B^{(1)}_n,\zeta^{(1)}_n\bigr) \bigr),
\end{eqnarray*}
where the sequences $\{\zeta^{(1)}_n\},\{S^{(1)}_{1i}\}, \{
S^{(1)}_{2i}\}, \ldots$ are defined in the same way
as $\{\zeta_n\}, \{S_{1i}\}, \{S_{2i}\}, \ldots$ introduced in
Section~\ref{sec.mod.sli}.

The Markov chain $\{\{S_n^{(z)}\}, z\ge1\}$ of Mat{\'e}rn processes
introduced above can be seen as a stationary
point process in 3D. It possesses
seven (free) parameters: $\lambda_c,\lambda_d,
a,b$ describing its initial distribution, and
$p,r^\prime,r^{\prime\prime}$ describing the transitions from step
to step, whereas the ``birth intensity'' $\lambda_c^\prime$ of (new)
cluster centers is given by $\lambda_c^\prime=\lambda_c (1-p)$.
It turned out that suitable choices for
$r^\prime,r^{\prime\prime}$ are the values of
$r^\prime=\sqrt{2}/2$ and $r^{\prime\prime}=1.5$. This means that
the uniform distribution of the displacement vectors
$D_1^{(z)},D_2^{(z)},\ldots$ is implemented as (discrete) uniform
distribution on the $8$-neighborhood
in the considered slice of voxels. Techniques for fitting the remaining
five parameters of the Markov chain $\{\{S_n^{(z)}\}, z\ge1\}$
are discussed in Section~\ref{sub.mod.mod}.

\subsubsection{Modeling the radii of spheres}\label{sec.Sim.rad}

\begin{figure}[b]

\includegraphics{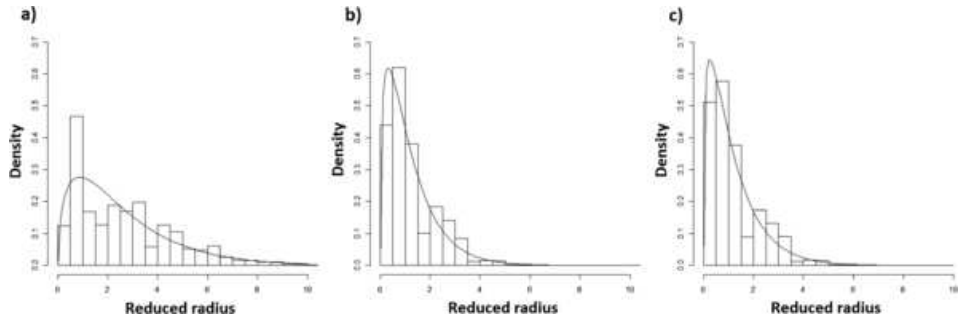}

\caption{Histograms of reduced radii and fitted gamma distributions (solid
lines); $\mathrm{a} = 57$~nm, $\mathrm{b} = 100$ nm, $\mathrm{c} = 167$ nm.}\label{fig.gamma_distribution}
\end{figure}

To get an idea which class of mark distributions is suitable to
model the radii of spheres, we computed histograms of radii which
have been extracted from the ET images for each of the three
photoactive layers with thicknesses of 57~nm, 100~nm and 167~nm.
Recall that in the sphere-putting algorithm mentioned in
Section~\ref{sec.multi-scale} we only consider spheres with a
minimum radius of $\sqrt{3}$ voxel sizes. Hence, instead of
computing histograms for the original radii, say, $r_1,r_2,\ldots,$
we computed histograms for correspondingly reduced radii
$r^\prime_1,r^\prime_2,\ldots,$ where $r_n^\prime=r_n-\sqrt{3}$.
It turns out that for all three film thicknesses, gamma
distributions can be fitted quite nicely to the histograms of
reduced radii; see Figure~\ref{fig.gamma_distribution}. The
parameters $k$ and $\theta$ of these gamma distributions
$\Gamma(k,\theta)$ have been estimated using the method of
moments; see Table~\ref{tab.gamma_distribution}.

\begin{table}[t]
\caption{ Parameters for gamma
distributions of radii}\label{tab.gamma_distribution}
\begin{tabular}{@{}lccc@{}}
\hline
\textbf{Parameter} & \textbf{57~nm film} & \textbf{100~nm film} & \textbf{167~nm film}\\
\hline
$k$ & 1.51 & 1.36 & 1.26 \\
$\theta$ & 1.73 & 0.88 & 0.93 \\
\hline
\end{tabular}
\end{table}

\begin{figure}[b]

\includegraphics{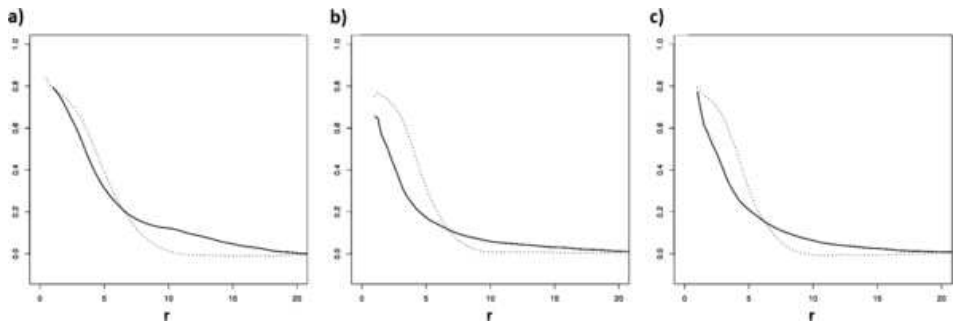}

\caption{Comparison of estimated (solid) and simulated (dotted) mark
correlation functions for arbitrary distance vectors in 3D of
length $r$; $\mathrm{a} = 57$~nm, $\mathrm{b} = 100$ nm, $\mathrm{c} = 167$~nm.}
\label{fig.mar.cor.sim}
\end{figure}

The mark correlation function of stationary marked
point processes is considered, which describes the spatial
correlations of pairs of marks, depending on the distance vector
of the corresponding pairs of points; see, for example, \citet{Illian08}.
For each representation of the three photoactive layers by unions
of overlapping spheres, the values $\widehat\kappa(r)$ of this
function have been estimated for distance vectors of length $r$.
They show strong positive correlations for radii corresponding to
pairs of midpoints with small distances from each other; see
Figure~\ref{fig.mar.cor.sim}.

Thus, for a given configuration $\{s_n^{(z)}, n, z\ge1\}$ of
midpoints, the radii $\{R_n^{(z)}, n, z\ge1\}$ associated with
these midpoints are not modeled just by independent marking, but
the following moving-average procedure is proposed. For some
$m\ge1$, let $\{\widetilde R_n^{(z)}, n,z\ge1\}$ be an i.i.d.
sequence of $\Gamma(k/m,\theta)$-distributed random variables, and
let $(z_1,n_1),\ldots,(z_m,n_m)$ for each index $(z,n)$ denote
the indices of the $m$ nearest neighbors
$s_{n_1}^{(z_1)},\ldots,s_{n_m}^{(z_m)}$ of $s_{n}^{(z)}$
[including the point $s_{n}^{(z)}$ itself]. Then, the radius
$R_n^{(z)}=\sqrt{3}+\widetilde R_{n_1}^{(z_1)}+\cdots+\widetilde
R_{n_m}^{(z_m)}$ is assigned to the midpoint $s_n^{(z)}$. The
reduced radius $R_n^{(z)}-\sqrt{3}$ obtained in this way is
$\Gamma(k,\theta)$-distributed. It turned out that for $m=4$, the
estimated mark correlation functions computed from real data (i.e.,
original point pattern and original radii) show a good resemblance
to their simulated counterparts (i.e., original point pattern and
simulated radii) for all three thicknesses of photoactive layers;
see Figure~\ref{fig.mar.cor.sim}.

\subsubsection{Model fitting for midpoints of spheres}\label{sub.mod.mod}

The (overall) intensity of midpoints of spheres, that is, the intensity
$\lambda$ of the stationary point process
$\{S_n^{(z)}, z, n\ge1\}$ can be easily estimated by $\widehat\lambda=\#
\{S_n^{(z)}\dvtx S_n^{(z)}\in W\}/|W|$, where
$\#\{S_n^{(z)}\dvtx S_n^{(z)}\in W\}$ is the total number of midpoints
in the sampling window $W\subset\mathbb{R}^3$
and $|W|$ denotes the volume of $W$. Using this formula, the following
values have been obtained for the spheres extracted from the binarized
ET images:
$\widehat\lambda= 1.83 \cdot10^{-3}$ for the 57~nm
film, $\widehat\lambda= 5.29 \cdot10^{-3}$ for the 100~nm film, and
$\widehat\lambda= 5.15 \cdot10^{-3}$ for the 167~nm film.

Note that $\lambda=\lambda_c \lambda_d |\mathcal{E}_{a,b}|$,\vspace*{1pt}
where $|\mathcal{E}_{a,b}|$ denotes the area of an ellipse with
semi-axes $a$ and $b$. Therefore, in order to determine $\lambda_d$,
the estimator
$\widehat\lambda_d=\widehat\lambda(\widehat\lambda_c|\mathcal
{E}_{\widehat
a,\widehat b}|)^{-1}$ can be used, provided that an
estimator $\widehat\lambda_c$\vspace*{-2pt} for the intensity~$\lambda_c$ of cluster
centers as well as estimators $\widehat a$ and~$\widehat b$ for
the semi-axes $a$ and~$b$ are available. Similarly to
the estimation of $\lambda_d$ discussed above, the estimator
$\widehat\lambda_c^\prime=\widehat\lambda_c(1-\widehat p)$ for the
birth rate $\lambda_c^\prime$ can be considered, provided that
estimators $\widehat\lambda_c$ and $\widehat p$ for $\lambda_c$
and $p$ are given.

Finally, we derive a so-called minimum-contrast estimator for the
vector of the remaining four parameters $\lambda_c$, $a$, $b$ and
$p$, where we traverse the parameter space of these parameters.
This means that for each vertex of a certain lattice of parameter
vectors $(\lambda_c, a,b,p)$, the 3D point process
$\{S_n^{(z)}, z,n\ge1\}$ of midpoints described in
Sections~\ref{sec.mod.sli} and \ref{sec.spa.bir} is simulated in
the sampling window $W$ and gamma-distributed radii are added
according to the moving average procedure described in
Section~\ref{sec.Sim.rad}. Then, structural characteristics of
simulated unions of spheres are compared with corresponding
structural characteristics of unions of spheres extracted from
binarized ET images.

In particular, consider the empirical distribution function
$\widehat F^{\mathit{SCD}}\dvtx[0,\infty)\to[0,1]$ of spherical contact
distances from polymer to ZnO,\vspace*{1pt} computed for the macro-scale
component of binarized ET images, and let $\widehat
F^{(x)},\widehat F^{(y)},\widehat F^{(z)}\dvtx\break[0,\infty)\to[0,1]$
denote the empirical chord-length distribution functions of the
ZnO domains in these images along the $x$-, $y$- and $z$-axis,
respectively. More information on spherical contact distance
distributions and chord length distributions can be found in
\citet{OHS00} and in \citet{Stoyan95}.\vspace*{1pt}
Consider the volume fraction $\widehat
V$ of ZnO, computed for the macro-scale component of binarized ET
images, and let $\widehat{V}{}^{\prime}$ denote the volume fraction
of those ZnO voxels of the macro-scale component contributing to percolation
pathways (monotonous and nonmonotonous) through the photoactive layer.

Let
$F^{\mathit{SCD}}_{\lambda_c, a, b,p}$, $F^{(x)}_{\lambda_c, a, b,p}$,
$F^{(y)}_{\lambda_c, a, b, p}$, $F^{(z)}_{\lambda_c, a,
b,p}$
be the corresponding distribution
functions and $V_{\lambda_c, a,
b,p}$, $V^\prime_{\lambda_c, a,
b,p}$ the volume fractions, respectively,\vspace*{1pt} obtained from
simulated 3D morphologies in dependency of $\lambda_c$, $a$, $b$
and $p$. Then, each solution $(\widehat\lambda_c, \widehat a,
\widehat b,\widehat p)$
of the minimization problem
\begin{eqnarray*}
(\widehat\lambda_c, \widehat a, \widehat b,\widehat p)
&=& \operatorname{argmin}\limits_{\lambda_c,a,b,p} \bigl( w^{\mathit{SCD}} \|
\widehat F^{\mathit{SCD}}-{F}^{\mathit{SCD}}_{\lambda_c,a,b,p}
\|
+ w^{(x)} \bigl\| \widehat F^{(x)}-F^{(x)}_{\lambda_c,a,b,p} \bigr\| \\
&&\hphantom{\operatorname{argmin}\limits_{\lambda_c,a,b,p} \bigl( }
{}+ w^{(y)} \bigl\| \widehat F^{(y)}-F^{(y)}_{\lambda_c,a,b,p} \bigr\|
+ w^{(z)} \bigl\| \widehat F^{(z)}-F^{(z)}_{\lambda_c,a,b,p}\bigr\|\\
&&\hphantom{\operatorname{argmin}\limits_{\lambda_c,a,b,p} \bigl( }\hspace*{20pt}
{}+ w^{(V)} |\widehat V-V_{\lambda_c, a,
b,p} |
+ w^{(V^\prime)} |\widehat{V}{}^{\prime}-V^\prime_{\lambda_c, a,
b,p} |
\bigr)
\end{eqnarray*}
is called a minimum-contrast estimator for $(\lambda_c, a, b,p)$,
where  $w^{\mathit{SCD}},w^{(x)},w^{(y)},\break w^{(z)}\ge0$ and
$w^{(V)},w^{(V^\prime)}\ge0$ are some weights such that
$w^{\mathit{SCD}}+w^{(x)}+w^{(y)}+w^{(z)}w^{(V)}+w^{(V^\prime)}=1$, and
$\|
\widehat F-F\|= \sup_{t \in\mathbb{R}} | \widehat F(t)-F(t)|$
denotes the Kolmogorov distance of $\widehat F$ and $F$.

The minimization problem described above is solved numerically,
that is, only a relatively coarse lattice of parameter vectors
$(\lambda_c, a, b, p)$ can be taken into account.
Table~\ref{tab.paraters_pointprocess} summarizes the results
obtained in this way, where we put $w^{\mathit{SCD}}=
w^{(V)}=w^{(V^\prime)}=1/4$ and $w^{(x)}=w^{(y)}=w^{(z)}=1/(3\cdot
4)=1/12$. The results shown in
Table~\ref{tab.paraters_pointprocess} nicely reflect the main
structural differences and similarities of the patterns of sphere
midpoints for the 57 nm, 100 nm, and 167 nm films: The estimated
values for $\lambda_c$, $a$ and $b$ indicate that the 57 nm film
has fewer, but larger clusters of midpoints than the 100 nm and
167 nm films, whereas the intensity $\lambda_d$ of cluster members
is similar for all three films; see also Figure~\ref{fig.cor.fun}.
The survival probability $p$ is close to $1$ and, therefore, the
birth rate $\lambda_c^\prime$ is much smaller than the intensity
$\lambda_c$ of ``old'' cluster centers; see
Figure~\ref{fig.SuccessiveMidpoints}.

\begin{table}
\caption{Parameters for 3D point processes of midpoints}\label{tab.paraters_pointprocess}
\begin{tabular}{@{}lccc@{}}
\hline
\textbf{Parameter} & \textbf{57~nm film} & \textbf{100~nm film} & \textbf{167~nm film} \\
\hline
$\lambda_c $ & 9.0E$-$5 & 1.25E$-$3 & 1.00E$-$3 \\
$a$ & 45 & 22 & 24 \\
$b$ & 15 & 6 & 10 \\
$p$ & 0.987 & 0.991 & 0.977 \\
$\lambda_d$ & 9.59E$-$3 & 10.0E$-$3 & 6.83E$-$3 \\[1pt]
$\lambda_c^\prime$ & 1.17E$-$6 & 1.17E$-$5 & 2.34E$-$5 \\
\hline
\end{tabular}
\end{table}

\begin{figure}

\includegraphics{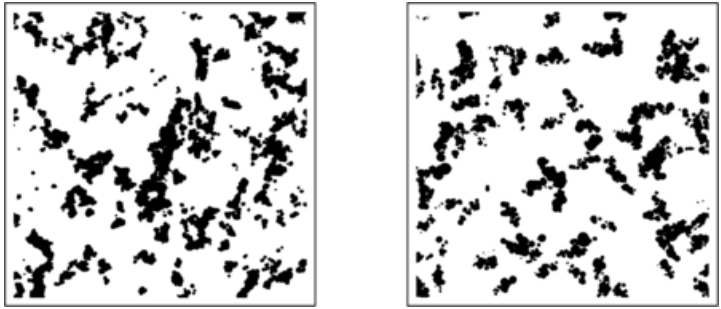}

\caption{Left: 2D slice of the morphologically smoothed 57~nm film,
right: 2D slice
of a~simulated union of overlapping spheres, drawn from the fitted
3D model.}\label{fig.RealisationMacroScale}
\end{figure}

We also remark that the method of statistical model-fitting
explained in this section leads to a relatively high degree of
visual coincidence between simulated and real ET images; see
Figure~\ref{fig.RealisationMacroScale}. A more formal approach to
model validation will be given later on in
Section~\ref{sec.mod.val}.

\subsection{Stochastic modeling of clusters of misspecified
voxels}\label{sec.SimulationModelMicro}

We now develop a modeling approach for the micro-scale part of
the morphological structure. It is used to stochastically invert
the morphological smoothing and completes our model for the 3D
morphology of hybrid polymer-ZnO solar cells. The micro-scale
morphology is modeled separately for each of the three types of
misspecifications, that is, the micro-scale components of outer,
boundary, and interior misspecifications mentioned in
Section~\ref{sec.multi-scale}.

\subsubsection{Outer misspecifications}\label{sub.out.mis}

Recall that each ZnO voxel that is not covered by a sphere, and
therefore constituted as polymer, is said to be an outer
misspecification. They typically form thin branches or small
isolated ZnO particles. In the present section, a stochastic model
is proposed for the locations and sizes of clusters of outer
misspecifications, that is, the connected components of the set
$B\setminus B^{\prime\prime\prime}$ introduced in
Section~\ref{sec.multi-scale}. We first consider this kind of
misspecification, because it influences the ``correction'' of
boundary misspecifications which will be described in
Section~\ref{sub.bou.mis} below.

We assume that the centers of gravity of clusters of outer
misspecifications form a Cox point process, which is also called a
doubly stochastic Poisson process in literature.
The cluster sizes are considered as marks. In particular, under the
condition that a realization $\{(s_n^{(z)},r_n^{(z)})\}$
of the (marked) point-process model $\{(S_n^{(z)},R_n^{(z)}),n,z\ge1\}
$ introduced
in Section~\ref{sec.SimulationModelMacro} is given which describes the
macro-scale component of the ZnO phase represented by the union of
spheres $\xi=\bigcup_{n,z\ge1} b( s_n^{(z)},r_n^{(z)})$, we assume that
the centers of gravity of clusters of outer
misspecifications can be described by an inhomogeneous Poisson process.
Its (conditional) intensity $\lambda(x)$ at location
$x\in\mathbb{R}^2$ depends
on the distance $\delta_\xi(x)=\inf\{|x-y|\dvtx y\in\xi\}$ between
$x$ and the union of spheres~$\xi$, where we put $\lambda(x)=0$ if
$\delta_\xi(x)=0$.
%

The intensity $\lambda(x)$ at locations $x\in\mathbb{R}^2$ with $\delta
_\xi(x)>0$ can be estimated by analyzing
the centers of gravity of clusters of outer
misspecifications in the set $B\setminus B^{\prime\prime\prime}$
extracted from binarized ET images. 
In particular, for any $d_l,d_u>0$ with $d_l<d_u$, the (average)
intensity $\lambda_{[d_l,d_u)}$ of centers of gravity
at locations $x\in\mathbb{R}^2$ with $\delta_\xi(x)\in[d_l,d_u)$ can be
estimated by
{\small
\[
\widehat{\lambda}_{[d_l,d_u)}=
\frac{\mathrm{number\ of\ centers\ of\ gravity\ with\ distance\ to\ }
B^{\prime\prime\prime}\mathrm{\ between\ }d_l\ \mathrm{and}\ d_u}
{\mathrm{number\ of\ voxels\ with\ distance\ to}\ B^{\prime\prime\prime
}\ \mathrm{between}\ d_l\ \mathrm{and}\ d_u} .
\]}
Examples of results for $\widehat{\lambda}_{[d_l,d_u)}$ computed from
ET images are given in Table~\ref{tab.outer_misspecifications}.
For all three film thicknesses the estimated intensity $\widehat{\lambda
}_{[d_l,d_i)}$
decreases with increasing distance to
the macro-scale component $B^{\prime\prime\prime}$ of the ZnO phase.

\begin{table}
\caption{Model parameters for
outer misspecifications}\label{tab.outer_misspecifications}
\begin{tabular}{@{}lcccc@{}}
\hline
& & \multicolumn{1}{c}{\textbf{57~nm}} & \multicolumn{1}{c}{\textbf{100~nm}} & \multicolumn{1}{c@{}}{\textbf{167~nm}} \\
\hline
Intensity & $\lambda_{[0,2)}$ & \multicolumn{1}{c}{1.78E$-$3} & \multicolumn{1}{c}{5.25E$-$3} & \multicolumn{1}{c@{}}{5.70E$-$3} \\
& $\lambda_{[2,4)}$ & \multicolumn{1}{c}{5.22E$-$4} & \multicolumn{1}{c}{4.31E$-$4} & \multicolumn{1}{c@{}}{6.88E$-$4} \\
& $\lambda_{[4,6)}$ & \multicolumn{1}{c}{1.81E$-$4} & \multicolumn{1}{c}{1.80E$-$4} & \multicolumn{1}{c@{}}{4.48E$-$4} \\
& $\lambda_{[6,8)}$ & \multicolumn{1}{c}{1.04E$-$4} & \multicolumn{1}{c}{1.10E$-$4} & \multicolumn{1}{c@{}}{3.28E$-$4} \\
& $\lambda_{[8,10)}$ & \multicolumn{1}{c}{6.90E$-$5} & \multicolumn{1}{c}{6.06E$-$5} & \multicolumn{1}{c@{}}{2.31E$-$4} \\[3pt]
Slope & $\alpha$ & $-0.90$ & $-0.67$ & $-1.13$ \\
Axis intercept & $\beta$ & 86.45 & 33.62 & 30.92 \\
Variance & $\sigma^2$ & 1,889.5 & 114.6 & 63.9 \\
\hline
\end{tabular}
\end{table}

\begin{figure}[b]

\includegraphics{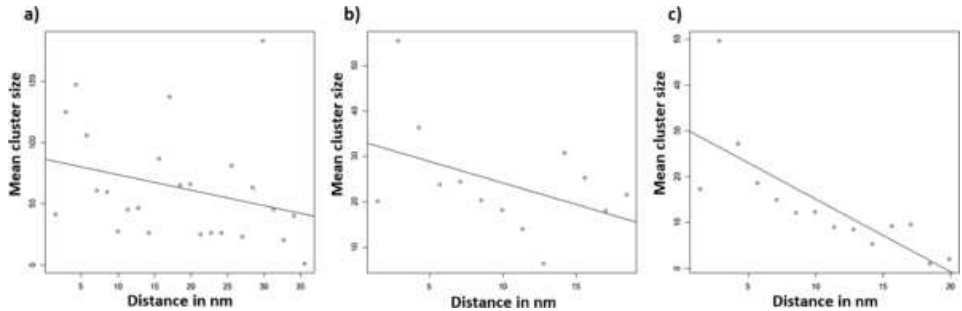}

\caption{Mean cluster sizes of outer misspecifications, depending on their
distance from the set $B^{\prime\prime\prime}$;
$\mathrm{a} = 57$~nm, $\mathrm{b} = 100$ nm, $\mathrm{c} = 167$ nm.}\label{fig.OuterMisSize}
\end{figure}

We assume that within shells around the set $\xi=\bigcup_{n,z\ge1}
b( s_n^{(z)},r_n^{(z)})$ with distance to $\xi$ in the distance
class $[d_l, d_u)$, the intensity of centers of gravity of outer
misspecifications is constant and given by $\lambda_{[d_l,d_u)}$.
We assume that the clusters of outer misspecifications are spheres,
the radii of which are given in the following way. It turned out
that not only the intensity of centers of gravity, but also the
cluster sizes observed in binarized ET images, depend on the
distances of centers of gravity to the set
$B^{\prime\prime\prime}$. In particular, the cluster sizes seem to
have a tendency to decrease with increasing distance to the set
$B^{\prime\prime\prime}$; see Figure~\ref{fig.OuterMisSize}, where
the estimated mean values for the radii of the considered distance
classes are shown.

To integrate this dependency into our simulation model,
we fit regression lines to the point clouds shown in
Figure~\ref{fig.OuterMisSize}, that is, we assume that the
points in this figure can be seen as realizations of
random variables $Y_i$ satisfying the linear relation
$Y_i = \alpha x_i + \beta+ \varepsilon_i$,
where $x_i = (d_l^{(i)} + d_u^{(i)})/{2}$ is the midpoint of the
$i$th distance
class $[ d_l^{(i)}, d_u^{(i)})$ and $\varepsilon_i$ is a random error term.
The parameters $\alpha$ and $\beta$ of this regression line are
estimated by the method of least
squares. 
As can be seen in
the plots of Figure \ref{fig.OuterMisSize}, a linear model
shows just the trend but is not a perfectly fitting model
for describing the cluster sizes in dependence on their distances to
the macro-scale component~$B^{\prime\prime\prime}$ in the ZnO phase.
Hence, we additionally consider the residua~$\varepsilon_i$ in the linear
regression model, where we assume that they follow a
normal distribution with expectation $0$ and variance $\sigma^2$.
The estimated values obtained for slope~$\alpha$, intercept $\beta$ of $y$-axis,
and variance $\sigma^2$ of the residua are given in
Table~\ref{tab.outer_misspecifications}.
To ensure a positive size of each simulated cluster, we reject
negative sizes and generate new realizations as long as a positive
cluster size is sampled.
For simulated
clusters of outer misspecifications with a greater distance
from the set $\xi$ than the intercept of the fitted regression
line with the $x$-axis, we put the cluster size equal to zero.
This is in accordance with real data, because in the
binarized ET images such clusters of outer misspecifications
do not occur. In the following, by $\xi^\prime$ we will denote a realization
of the model with included outer misspecifications.

\subsubsection{Boundary misspecifications}\label{sub.bou.mis}

After adding the outer misspecifications to our model as described
in the previous section, we now develop an algorithm to remove the
so-called boundary misspecifications, which primarily result from
the dilation of the ZnO domains; see Section
\ref{sec.multi-scale}. Recall that we defined the boundary
misspecifications as those misspecified voxels within some outer
shells of the set $B^{\prime\prime\prime}$, that is, the union of
spheres representing the morphologically smoothed ZnO phase. The
first shell is defined as the set of voxels of
$B^{\prime\prime\prime}$, with a distance to $W \setminus
B^{\prime\prime\prime}$ smaller or equal than~$1$. In general, the
$(i+1)$th shell, $i=1,2,3,\ldots,$ is defined as the set of voxels
of $B^{\prime\prime\prime}$ with a distance to $W \setminus
B^{\prime\prime\prime}$ in $(i,i+1]$.

Table \ref{tab.boundary_misspecifications} displays the percentage
of misspecified boundary voxels in the different outer shells of
$B^{\prime\prime\prime}$.
Note that boundary misspecifications only occur in the first outer shell
(for the 100 nm and 167 nm films) and in the
first three outer shells (for the 57 nm film), respectively.
As a consequence, the simulation model
should remove about the same percentage of boundary voxels in the
corresponding outer shells of $\xi$.

\begin{table}
\caption{Fractions of boundary
misspecifications
in consecutive shells (given in \%)}\label{tab.boundary_misspecifications}
\begin{tabular}{@{}lcccccc@{}}
\hline
& \multicolumn{2}{c}{\textbf{57~nm}} & \multicolumn{2}{c}{\textbf{100~nm}} & \multicolumn{2}{c@{}}{\textbf{167~nm}} \\[-5pt]
&\multicolumn{2}{c}{\hrulefill}&\multicolumn{2}{c}{\hrulefill}&\multicolumn{2}{c@{}}{\hrulefill}\\
& \textbf{ET data} & \textbf{Simulation} & \textbf{ET data} & \textbf{Simulation} & \textbf{ET data} & \textbf{Simulation} \\
\hline
1st shell & 87 & 72 & 60 & 67 & 60 & 64 \\
2nd shell & 74 & 70 & \hphantom{0}0 & \hphantom{0}0 & \hphantom{0}0 & \hphantom{0}0 \\
3rd shell & 71 & 68 & \hphantom{0}0 & \hphantom{0}0 & \hphantom{0}0 & \hphantom{0}0 \\
4th shell & \hphantom{0}0 & \hphantom{0}0 & \hphantom{0}0 & \hphantom{0}0 & \hphantom{0}0 & \hphantom{0}0 \\
\hline
\end{tabular}
\end{table}

Some parts of the outer shells of $B^{\prime\prime\prime}$ belong to
thin branches of ZnO, therefore not the complete shells of $\xi$
have to be removed.
To include such thin branches into the model, we combine the
model for the outer misspecifications introduced in Section
\ref{sub.out.mis} with the following
algorithm to remove the boundary misspecifications.

For a given realization $\xi$,
we iteratively remove those parts of the
outer shells of $\xi$ which are not connected to the set $\xi^{\prime}
\setminus\xi$
introduced in the previous section.
In more detail, to correct the first outer shell, we first
determine the set of all voxels $\eta\subset\xi$ belonging to the
first outer shell.
Subsequently, all voxels of the set $\eta_1 \subset\eta$ that are not touching
the set $\xi^{\prime} \setminus\xi$, that is, whose distance to $\xi
^{\prime} \setminus\xi$
is greater than 1, are removed. Now, the first outer shell is corrected.
Those parts $\eta_2= \eta\setminus\eta_1$ of the first outer shell
that have
not been removed since they were located near an outer misspecification
are---for technical reasons---added to the set $\xi^{\prime} \setminus
\xi$
of simulated outer misspecifications.
Hence, when correcting the second outer shell, the voxels near the
set $ (\xi^{\prime} \setminus\xi ) \cup\eta_2$ are not removed.
This reproduces the thin branches as observed in the binarized ET data.
To correct the third shell, the same procedure is repeated.
The result after additionally adding the boundary misspecifications
into the model is denoted by $\xi^{\prime\prime}$.

\subsubsection{Interior misspecifications}\label{sub.int.mis}

As mentioned in Section \ref{sec.multi-scale}, the remaining
misspecified voxels in $B^{\prime\prime\prime}$ are classified as
interior misspecifications. These interior misspecifications
typically form small polymeric holes
inside the ZnO domains.

Our modeling approach for the interior misspecifications is based
on the assumption that the polymeric holes in the ZnO domains
possess spherical shapes and are not overlapping, that is,
we consider them as hard spheres.

Similarly to the modeling of the outer misspecifications,
we assume that the centers of gravity of the interior
misspecification clusters form a doubly stochastic
point process, where again the cluster
sizes are considered as marks. We assume the
points of this point process
to have a certain minimum distance $r_h$ to each other
because the interior misspecifications are seen as
nonoverlapping spheres.
In particular, given the realization
$\xi^{\prime\prime}$, that is, a realization of the marked point
process for the macro-scale component of the ZnO phase
introduced in Section \ref{sec.SimulationModelMacro} with
included outer and boundary misspecifications as described in Sections
\ref{sub.out.mis} and \ref{sub.bou.mis},
the centers of gravity of interior misspecification clusters are
assumed to
form a (conditional) Mat\'ern hard-core process in $\xi\cap\xi^{\prime
\prime}$.
We assume that the marks of this point process
are spheres with a constant radius $r=\frac{r_h}{2}$,
that is, the interior misspecifications are modeled by nonoverlapping
spheres with equal radii.

The Mat\'ern hard-core process in $\mathbb{R}^3$ with intensity $\lambda_h$
and hard-core radius $r_h$ is a thinned homogeneous Poisson
point process, where the remaining points have a distance of
at least $r_h$ to each other.
Further details can be found, for example, in \citet{Illian08}.
Given the set $\xi\cap\xi^{\prime\prime}$, the centers of gravity of interior
misspecifications are then modeled by those points of the Mat\'ern hard-core
process which belong to $\xi\cap\xi^{\prime\prime}$.
Hence, this model for the
interior misspecifications can be called a doubly stochastic
Mat\'ern hard-core process.

\begin{table}
\caption{Estimated parameters for interior misspecifications}\label{tab.Interior}
\begin{tabular}{@{}lccc@{}}
\hline
\textbf{Parameter} & \textbf{57~nm film} & \textbf{100~nm film} & \textbf{167~nm film}\\
\hline
Radius $r$ & 2.50 & 1.30 & 1.07 \\
Intensity $\lambda_h$ & 1.37E$-$3 & 5.17E$-$3 & 5.12E$-$3 \\
\hline
\end{tabular}
\end{table}

To fit this model to real data, we first estimate the mean
volume $\overline{V}$ of the clusters for all three film thicknesses
and transform this (mean) volume into a radius $\hat{r}$ of a
ball with the same volume $\overline{V}$.
The corresponding radii obtained for the three considered data sets
can be seen in Table \ref{tab.Interior}. The hard-core radius $\hat
{r}_h$ of the
Mat\'ern hard-core process is then computed as $\hat{r}_h = 2\hat
{r}$, which ensures
that the spheres of radius $\hat{r}$ centered at the points of the
doubly stochastic Mat\'ern hard-core process
are not overlapping. For the intensity $\lambda_h$ of the Mat\'ern
hard-core process
we consider the following natural estimator,
\[
\widehat{\lambda}_h = \frac{ \mathrm{number\ of\ disjoint\ clusters\
of\ interior\ misspecifications}}
{|B^{\prime\prime} \ominus b(o,r)|},
\]
where $|B^{\prime\prime}
\ominus b(o,r)|$ denotes the volume of the set $B^{\prime\prime}
\ominus b(o,r)$. The values of $\widehat{\lambda}_h$ obtained for
the three data sets are shown in Table \ref{tab.Interior}. From
the results given in Table \ref{tab.Interior} it can be seen that,
also with respect to interior misspecifications, the 57 nm film
behaves rather different than the 100 nm, and 167 nm films: in the
first case, there are fewer, but larger clusters of interior
misspecifications than in the latter case. In the following, by
$\xi^{\prime\prime\prime}$ we denote a realization of our
simulation model after including all three types of
misspecifications.

\begin{figure}

\includegraphics{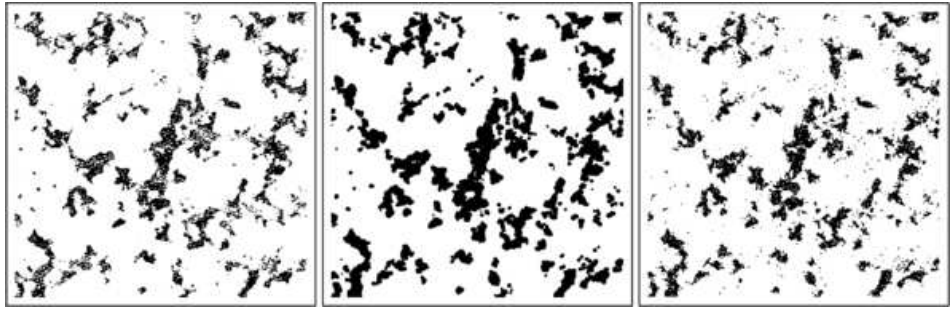}

\caption{Left: 2D slice of 57~nm film, center: corresponding
representation by a union of spheres, right: simulated correction
using the model for the misspecified voxels applied to the
representation by a union of
spheres.}\label{fig.RealisationInversionOfSmoothing}
\end{figure}

Figure \ref{fig.RealisationInversionOfSmoothing} shows a
realization of the model for the micro-structure, where
the micro-structure model is applied to the real data, more precisely,
the representation by a union of spheres $B^{\prime\prime\prime}$ of
the 57~nm film.
The two images on the left and the right sides of Figure
\ref{fig.RealisationInversionOfSmoothing} possess a high
degree of visual resemblance. See also Section
\ref{sec.mod.val} for a more formal approach to model validation.

\section{Model validation}\label{sec.mod.val}

To evaluate the goodness of fit, we compare model
characteristics which have been computed from real
and simulated data, respectively. On the one hand, we consider
structural characteristics of the ZnO nanomorphology
like the volume fraction of ZnO,
the volume fraction of ZnO contributing to monotonous
percolation pathways, and the distribution of spherical contact
distances from polymer to ZnO. On the other hand, we consider
a physical characteristic, the so-called exciton quenching
probability. This characteristic describes the probability
that a photo-excited particle generates charges. These
characteristics have also been used in \citet{Oosterhout09}
to characterize the morphology of a photoactive layer,
since they are closely related to the performance of solar cells.
To compare the values of these characteristics, obtained
from simulated and real data,
we binarize the ET images
using two extreme global thresholds as suggested in \citet{Oosterhout09}.
Recall that the two global thresholds have been chosen in such a way
that the
ZnO phase can be assumed to be a subset of the union of
foreground voxels (high threshold) and,
vice versa, the polymer phase is contained in
the union of background voxels (low threshold).

\begin{figure}

\includegraphics{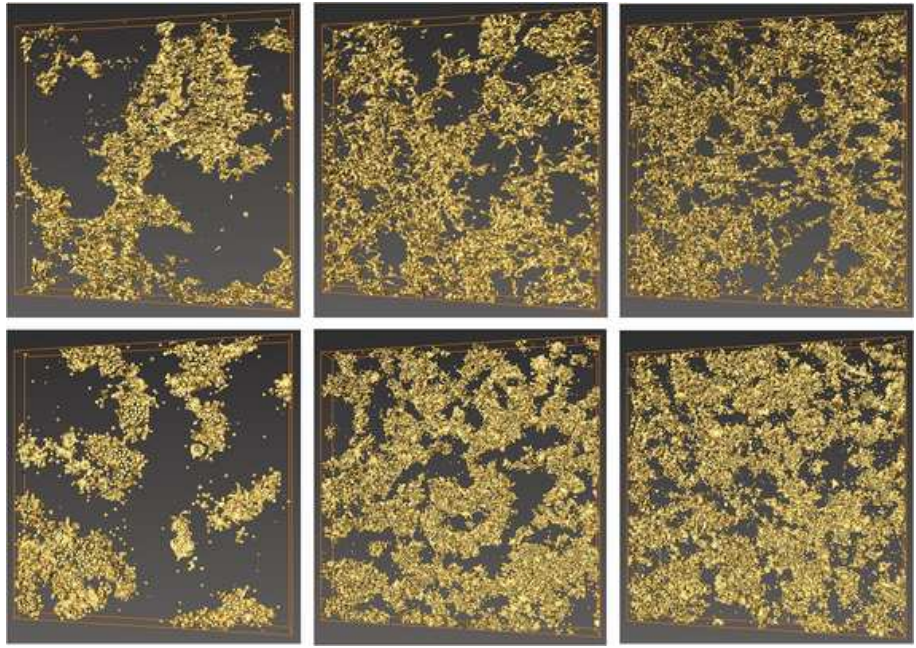}

\caption{3D cutouts ($400 \times400 \times37 $ voxels) of binarized
ET images obtained by adaptive thresholding (1st row) and realizations
of the complete model (2nd row); left: 57~nm film, center: 100~nm
film, right: 167~nm film.}\label{fig.RealisationsComplete}
\end{figure}

It turns out that the estimated values obtained for most of the considered
image characteristics of these two binarizations can be seen as
lower and upper bounds, respectively, for corresponding values
obtained for simulated images.
In addition to this,
we mention that, in accordance with the visual resemblance
of images obtained from real and simulated data for the
macro-scale component (see Figure \ref{fig.RealisationMacroScale}) and
the micro-scale
component (see Figure \ref{fig.RealisationInversionOfSmoothing}) of the
ZnO nanomorphology,
the optical resemblance between binarized ET images obtained
by adaptive thresholding and realizations of the complete
simulation model is also quite well; see Figure~\ref{fig.RealisationsComplete}.

\subsection{Checking morphological characteristics}\label{sub.mor.cha}

For a quantitative validation of the stochastic simulation model,
we first consider structural characteristics of the ZnO
nanomorphology.
For this purpose, we generate $100$ realizations of our model,
estimate the considered characteristics for each of these realizations
and compute their mean
values. In the case of the spherical contact distance distribution function,
the pointwise means are considered.

First, the volume fraction of ZnO is considered, which is one of the
most important characteristics in structural modeling. The
results given in Table~\ref{tab.check_morphological}
show that for all three film thicknesses, the volume
fractions of ZnO computed from simulated data are between
the corresponding bounds obtained from the globally thresholded ET
images.

\begin{table}
\caption{Volume fractions of ZnO for
globally thresholded and simulated images}\label{tab.check_morphological}
\begin{tabular}{@{}lcccc@{}}
\hline
& && & \textbf{Volume fraction}\\
& &  \textbf{Volume fraction}  &  \textbf{Volume fraction}& \textbf{with monotonous}\\
& &\textbf{of ZnO} & \textbf{with connection} & \textbf{connection} \\
\hline
\hphantom{0}57 nm& Low threshold & 0.098 & 0.934 & 0.872 \\
& Simulated data & 0.112 & 0.905 & 0.864 \\
& High threshold & 0.172 & 0.974 & 0.947 \\[3pt]
100~nm & Low threshold & 0.133 & 0.890 & 0.673 \\
& Simulated data & 0.215 & 0.971 & 0.910 \\
& High threshold & 0.295 & 0.991 & 0.936 \\[3pt]
167~nm & Low threshold & 0.128 & 0.851 & 0.630 \\
& Simulated data & 0.210 & 0.943 & 0.806 \\
& High threshold & 0.293 & 0.979 & 0.907 \\
\hline
\end{tabular}
\end{table}

\begin{figure}[b]

\includegraphics{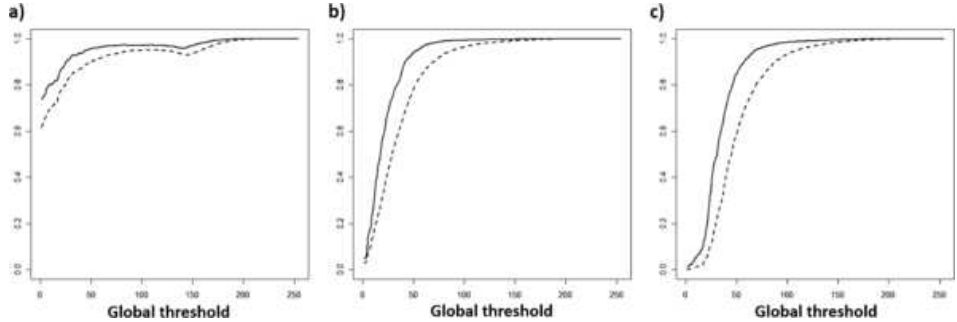}

\caption{Volume fractions of connected (solid lines) and monotonously connected
(dashed lines) foreground phase, in dependence of the global threshold;
$\mathrm{a} = 57$ nm, $\mathrm{b} = 100$ nm, $\mathrm{c} = 167$ nm.}\label{fig.connectivity}
\end{figure}

In the next step, the connectivity of the ZnO phase is considered.
This also is an important characteristic, because only if there is
a high connectivity, that is, if many percolation pathways exist,
the produced charges can be transported to the electrodes,
where current can be gripped. For estimating the connected
and monotonously connected volume fractions of ZnO we applied the same
methods as in \citet{Oosterhout09}.
However, in general, the (conditional) connected and
monotonously connected volume fractions of the foreground phase
in globally
thresholded images do not monotonously
depend on the values of global thresholds.
But as shown in Figure~\ref{fig.connectivity}, this is the
case for the globally thresholded ET data of photoactive layers of polymer-ZnO
solar cells.
Hence, also with respect to connected volume fractions,
the values for the two extreme thresholds can be
seen as upper and lower bounds.
With the exception of the 57~nm film, the values for the
connected volume fractions computed from simulated data,
given in Table \ref{tab.check_morphological}, are nicely
between the corresponding values obtained from globally
thresholded ET images. Also, the relative errors between the values
obtained for the adaptively thresholded ET images and simulated images
are rather small; see Table \ref{tab.rel.err.vol}.

\begin{table}
\caption{Volume fractions of ZnO for
adaptively thresholded and simulated images}\label{tab.rel.err.vol}
\begin{tabular}{@{}lcd{2.3}d{2.3}d{2.3}@{}}
\hline
& &&& \multicolumn{1}{c@{}}{\textbf{Volume fraction}}\\
& & \multicolumn{1}{c}{\textbf{Volume fraction}}  & \multicolumn{1}{c}{\textbf{Volume fraction}}  & \multicolumn{1}{c@{}}{\textbf{with monotonous}}\\
& & \multicolumn{1}{c}{\textbf{of ZnO}} &  \multicolumn{1}{c}{\textbf{with connection}}& \multicolumn{1}{c@{}}{\textbf{connection}} \\
\hline
\hphantom{0}57~nm & Adaptive threshold & 0.133 & 0.963 & 0.928 \\
& Simulated data & 0.112 & 0.905 & 0.864 \\
& Relative error & -0.158 & -0.060 & -0.069 \\[3pt]
100~nm & Adaptive threshold & 0.211 & 0.980 & 0.888 \\
& Simulated data & 0.215 & 0.971 & 0.910 \\
& Relative error & 0.019 & -0.009 & 0.025 \\[3pt]
167~nm & Adaptive threshold & 0.209 & 0.970 & 0.851 \\
& Simulated data & 0.210 & 0.943 & 0.806 \\
& Relative error & 0.005 & -0.028 & -0.053 \\
\hline
\end{tabular}
\end{table}

\begin{figure}[b]

\includegraphics{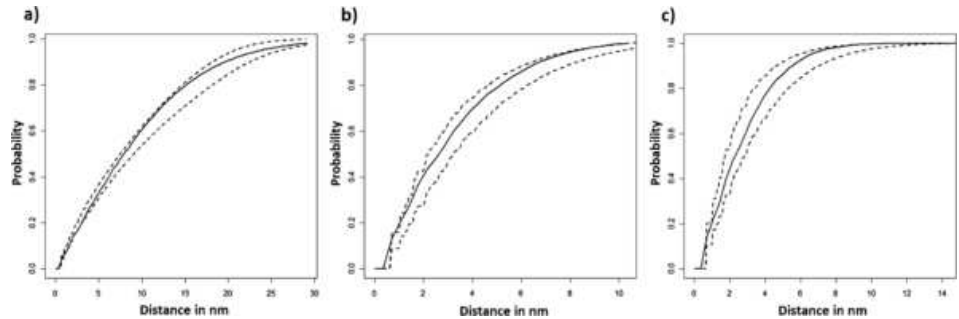}

\caption{Spherical contact distribution
functions. The lower and upper bounds
$\widehat{F}^{\mathit{SCD}}_l$ and $\widehat{F}^{\mathit{SCD}}_u$ obtained
from globally thresholded ET images
are plotted as dashed lines, the corresponding results from simulated
image data as solid lines;
$\mathrm{a} = 57$ nm, $\mathrm{b} = 100$ nm, $\mathrm{c} = 167$ nm.}\label{fig.Spherical}
\end{figure}

Finally, the spherical contact distribution function
$F^{\mathit{SCD}}\dvtx[0,\infty) \rightarrow[0,1]$ of the ZnO phase is
considered, where $F^{\mathit{SCD}}(t)$ can be interpreted as a~conditional
probability that the minimum distance from a randomly chosen
location to the ZnO phase is smaller or equal than $t\geq0$,
provided that the considered location belongs to the polymer
phase.

Similarly to the situation which we observed for the
connected volume fractions considered above,
it turns out that the spherical contact distribution functions
of the foreground phase in globally thresholded ET images
depend monotonously on the value of the global threshold.
Hence, the estimated contact distribution functions
$\widehat{F}^{\mathit{SCD}}_l$ and $\widehat{F}^{\mathit{SCD}}_u$ obtained for
the two
extreme thresholds can be seen as upper and lower bounds,
respectively; see Figure \ref{fig.Spherical}. In addition to this,
the spherical contact distribution $\widehat{F}^{\mathit{SCD}}$ obtained from
simulated data is shown in Figure \ref{fig.Spherical}, where
$\widehat{F}^{\mathit{SCD}}_l(t) \leq\widehat{F}^{\mathit{SCD}}(t) \leq\widehat
{F}^{\mathit{SCD}}_u(t)$
for all considered $t\geq0$ and for all three layer thicknesses.

In summary, we can conclude that our model fits very well
to real data regarding the considered structural characteristics.
As the model has been developed for analyzing the influence of
morphology on the performance of solar cells,
we also consider a physical characteristic for model validation
which is described in the following section.

\subsection{Checking probabilities of exciton quenching}\label{sub.pro.exi}

Quenching efficiency~$\eta_Q$ is the probability of a random
exciton being quenched; see Section \ref{sub.pho.act}. It is an
elementary but important physical characteristic for the
efficiency of solar cells.

In a hybrid polymer-ZnO solar cell, absorption
of light by the polymer phase does not directly
yield free charge carriers. Instead excitons
are formed. It is only at the interface of
the polymer and ZnO phase that free charges are
generated by quenching (splitting) of excitons.
It is therefore of the utmost importance that
excitons are able to reach this interface.
The exciton diffusion length in conjugated
polymers is typically a few nanometers, which puts
a considerable constraint on the morphology of
polymer solar cells. In other words, the efficiency
of exciton quenching is very sensitive to
morphology, making it a suitable way to
validate our model.

Suppose that the polymer phase $B^c = W \setminus B$
is given in a cubic sampling window $W\subset\mathbb{R}^3$.
Then, the overall efficiency $\eta_Q$ of exciton quenching
can be obtained from the field $ \{ n(x), x \in B^c \}$
of local exciton densities
in the polymer phase.
The exciton density field $ \{ n(x), x \in B^c \}$
can be computed by solving the steady-state
diffusion equation [see \citet{Oosterhout09}]
\[
0 = \frac{dn(x)}{dt} = - \frac{n(x)}{\tau}+D \nabla^2 n(x)+ g,\qquad  x \in B^c,
\]
where $D$ is the diffusion constant, $\tau$
is the exciton life time, and $g$ is the rate
of exciton generation.
As a boundary condition we require that all
excitons at the polymer-ZnO interface be quenched,
that is, $n(x) = 0 $ for all $ x\in\partial B^c \setminus
\partial W$.
For $x\in\partial B^c \cap\partial W$, cyclic boundary conditions
are applied in all directions. The exciton lifetime
and exciton diffusion rate in P3HT are taken from
the literature: $\tau = 400$ ps and
$D = 1.8 \cdot 10^{-7}$ m$^2$ s$^{-1}$; see \citet{Sha08}.
The rate of exciton generation $g$ is just a scaling factor,
where we use a value of $g=10^{27}$ m$^{-3}$ s$^{-1}$ which
is typical for 1-sun conditions.

%
\begin{figure}

\includegraphics{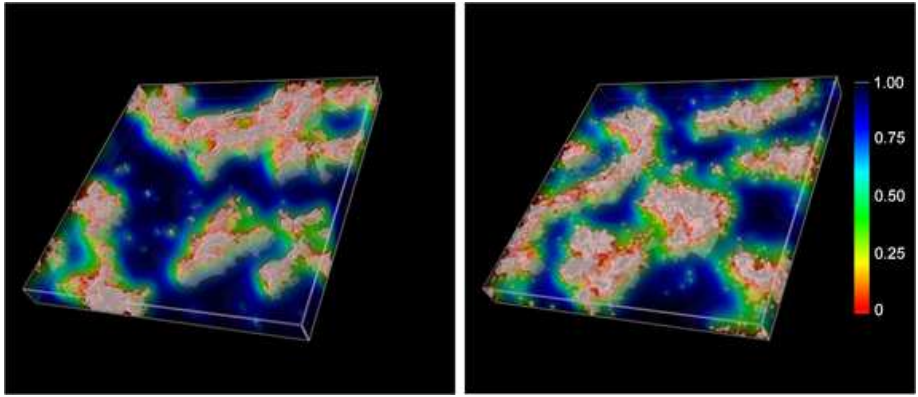}

\caption{The local exciton density $n(x)$ normalized
to $\tau g$ for adaptively thresholded ET (left) and simulated (right)
data. The scalebar, specifying the density of excitons,
applies to both images.}\label{fig.quenchingFields}
\end{figure}

The diffusion equation is solved numerically to a relative error
of less than 10$^{-3}$. Figure \ref{fig.quenchingFields} shows
local exciton density fields $ \{ n(x), x \in B^c \}$
for adaptively thresholded and simulated data, respectively.

\begin{figure}

\includegraphics{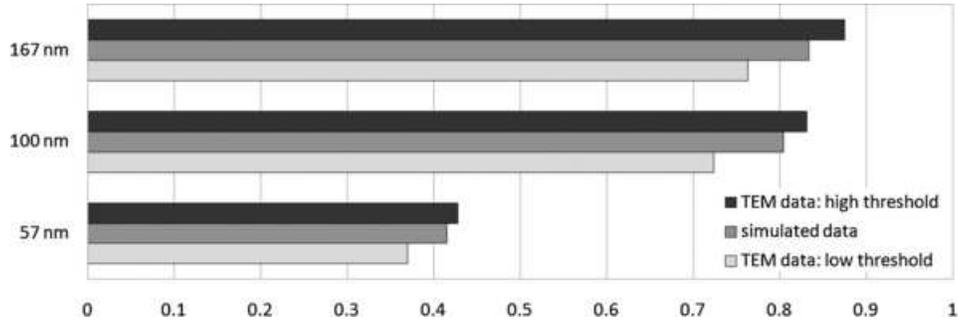}

\caption{Quenching efficiencies for globally
thresholded and simulated images.}\label{fig.quenching}
\end{figure}

Once $ \{ n(x), x \in B^c \}$ is known, the quenching efficiency
$\eta_Q$ follows from $\eta_Q = 1- \bar{n} /(\tau g)$,
where $\bar{n}$ is the average exciton density
in the polymer domain $B^c$. Figure \ref{fig.quenching}
compares the quenching efficiencies for original
and simulated data. The quenching
efficiency is also monotonously depending on the
global threshold. Hence, the values of $\eta_Q$ obtained for the two
extreme thresholds can be seen as lower and
upper bounds. The values of the quenching efficiency
for the simulated data lie well within these lower and
upper bounds; see Figure~\ref{fig.quenching}.
The small relative errors displayed in Table
\ref{tab.quenching} show that our model
fits very well to real data, where quenching efficiencies
of adaptively thresholded ET images are compared with those of
simulated data.

%

\begin{table}[b]
\tablewidth=261pt
\caption{Quenching efficiencies for adaptively
thresholded and simulated images}\label{tab.quenching}
\begin{tabular*}{\tablewidth}{@{\extracolsep{\fill}}ld{2.3}cc@{}}
\hline
& \multicolumn{1}{c}{\textbf{57 nm film}} & \textbf{100 nm film} & \textbf{167 nm film} \\
\hline
Adaptive threshold & 0.418 & 0.794 & 0.819 \\
Simulated data & 0.416 & 0.805 & 0.834 \\
Relative error & -0.010 & 0.014 & 0.018 \\
\hline
\end{tabular*}
\end{table}

\section{Conclusions and outlook}\label{sec.con.out}

In the present paper we developed a parameterized stochastic simulation
model for
the nanostructure of photoactive layers of hybrid polymer-ZnO solar cells.
The model is based on tools from stochastic geometry.
Additional to the model itself, we developed a method
to fit its parameters to real 3D ET image data.

To establish our model, the adaptively thresholded ET images are
segmented using a stochastic algorithm which consists of two main
steps. First, the images are morphologically smoothed in order to
slightly decrease their structural complexity. Then, the
morphologically smoothed binary images are represented by a system
of overlapping spheres, which can be interpreted as a realization
of a 3D marked point process, where the sphere centers are the
locations of points and the corresponding radii are their marks.
For the stochastic simulation model, we use a correlated vector of
2D elliptical Mat\'ern cluster processes, where the points are
subsequently marked to create a 3D marked point process. To
complete the model, that is, to include the structural details which
were omitted due to the morphological smoothing, a stochastic
simulation model for this ``micro-scale'' component is developed
afterward.

%

As our stochastic simulation model is fully parameterized,
we also developed techniques for
the estimation of the model parameters of all model components.
Thus, we are able to fit the simulation
model to the ET image data
described in Section \ref{sub.tom.gra}.

Finally, we validated the simulation model by comparing structural
and physical characteristics computed from simulated image data
with the corresponding characteristics obtained from globally and
adaptively thresholded ET images, respectively. In particular, the
quenching efficiencies computed for realizations of the simulation
model agree very well with those of the ET images. Hence, our
model nicely reflects the diffusion of excitons.

Since we were able to fit our model to 3D ET data, it
has already proved its capability to represent realistic
nanostructures of photoactive layers of hybrid polymer-ZnO solar cells.
The fact that the model is parameter-based enables
us to predict morphologies for film thicknesses, which have
not (yet) been imaged by 3D ET,
by interpolating or extrapolating the fitted model parameters.
Due to a strong correlation between
morphology and efficiency of polymer solar cells [see \citet
{Oosterhout09}],
the developed simulation model is of significant importance
for further investigations of polymer solar cells.
In a forthcoming paper we will
also investigate the transport processes of charges to the electrodes
as described in \citet{KOSTER10}, additional to the structural and
physical characteristics considered in the present paper.
By generating virtual morphologies, which are generated as realizations
of the developed model with different parameter configurations,
and investigating the transport processes of electrons and excitons therein,
the spatial stochastic model will be used
to identify morphologies of improved efficiency
with respect to the considered physical characteristics.

We also remark that the modeling approach developed in the present
paper can be applied to various other kinds of image data,
including geographical data considered, for example, in ecology. Then, for
instance, the Markov chain of Mat{\'e}rn cluster processes
introduced in Section~\ref{sec.spa.bir} may be viewed as a
sequence of dependent 2D point processes through time, which can
model, for example, the temporal movements of species in a given region.

\section*{Acknowledgments}
The authors would like to thank Image Analysis \& Stereology for
permission to reprint Figures \ref{fig.schematicView}--\ref{fig:MicroMacroScale} taken from \citet{TH10}.
Part of this work was done while Volker Schmidt was
visiting the Isaac Newton Institute for Mathematical Sciences,
Cambridge.

\printaddresses

\end{document}